\documentclass[twocolumn,superscriptaddress,showpacs,floatfix]{revtex4-2}

\UseRawInputEncoding
\usepackage{float}
\usepackage{dcolumn}
\usepackage{bm}
\usepackage[colorlinks = true, linkcolor = blue, urlcolor  = blue, citecolor = blue, anchorcolor = blue]{hyperref}
\usepackage{longtable}
\usepackage{booktabs}
\usepackage{mathrsfs}
\usepackage{graphicx,epsfig,latexsym,amssymb}
\usepackage{multirow,amsmath,array,booktabs,color}
\usepackage[section]{placeins}
\usepackage{color}
\usepackage{soul}

\begin{document}
\title{On the Origin of the Hidden Symmetry in the Asymmetric Quantum Rabi Model}
\author{Yun-Tong Yang}
\affiliation{School of Physical Science and Technology, Lanzhou University, Lanzhou 730000, China}
\affiliation{Lanzhou Center for Theoretical Physics, Key Laboratory of Theoretical Physics of Gansu Province, Key Laboratory of Quantum Theory and Applications of MoE, Gansu Provincial Research Center for Basic Disciplines of Quantum Physics, Lanzhou University, Lanzhou 730000, China}
\author{Song-Ming Chen}
\affiliation{School of Physical Science and Technology, Lanzhou University, Lanzhou 730000, China}
\affiliation{Lanzhou Center for Theoretical Physics, Key Laboratory of Theoretical Physics of Gansu Province, Key Laboratory of Quantum Theory and Applications of MoE, Gansu Provincial Research Center for Basic Disciplines of Quantum Physics, Lanzhou University, Lanzhou 730000, China}
\author{Hong-Gang Luo}
\email{luohg@lzu.edu.cn}
\affiliation{School of Physical Science and Technology, Lanzhou University, Lanzhou 730000, China}
\affiliation{Lanzhou Center for Theoretical Physics, Key Laboratory of Theoretical Physics of Gansu Province, Key Laboratory of Quantum Theory and Applications of MoE, Gansu Provincial Research Center for Basic Disciplines of Quantum Physics, Lanzhou University, Lanzhou 730000, China}

\begin{abstract}
The introduction of an asymmetric term into the quantum Rabi model generally lifts energy-level degeneracies. However, when the asymmetry parameter takes specific multiples of the bosonic mode frequency, level degeneracies reappear$-$a phenomenon referred to as the hidden symmetry in the asymmetric quantum Rabi model. Identifying the origin of this hidden symmetry and its explicit operator form constitutes two central tasks in studying this system. Here, we investigate the origin of this hidden symmetry using the method of two successive diagonalizations, with a focus on physics in the regime where the ratio between the two-level splitting $\Delta$ and the mode frequency $\omega$ satisfies $\Delta/\omega \gg 1$. We find that the hidden symmetry stems from energy-level matching within the asymmetric double-well potential, a picture strongly supported by the wavefunctions of both the ground and excited states. Moreover, the emergence of an excited-state quantum phase transition is identified and qualitatively discussed, which arises from the breaking and restoration of this hidden symmetry across different coupling regimes. Our results provide deeper insight into the physics of the asymmetric quantum Rabi model, particularly in the previously less-explored strong-coupling regime where $\Delta/\omega \gg 1$. 
\end{abstract}

\maketitle

\section{\label{sec:level1}Introduction}
The Rabi model was introduced by Rabi in 1936 to describe the interaction between a rapidly oscillating magnetic field and an atom with a fixed nuclear spin orientation, wherein the magnetic field is described by a classical electromagnetic wave, and the two-state spin system can be regarded as a two-level system \cite{Rabi1936, Rabi1937}. In 1963, Jaynes and Cummings proposed a quantized version, known as the Jaynes-Cummings (J-C) model \cite{Jaynes1963}, which incorporates the rotating-wave approximation on the quantum Rabi model (QRM), thereby ensuring the conservation of the total excitation number. This conserved quantity corresponds to a $U(1)$ symmetry, which renders the model exactly solvable \cite{Larson2021}. It was not until 2011 that the exact analytical solution of the full QRM was obtained, utilizing its discrete $\mathbb{Z}_2$ parity symmetry \cite{Braak2011, Chen2012}. Although the QRM appears formally simple, Hwang et al. discovered in 2015 that its ground state exhibits a superradiant phase transition (SPT) from the normal phase to the superradiant phase, accompanied by the breaking of the $\mathbb{Z}_2$ parity symmetry \cite{Hwang2015}. The universal scaling behavior and critical exponents indicate that this phase transition belongs to the Ising universality class \cite{Liu2017}. Subsequently, excited-state phase transitions in this model have also been investigated \cite{Puebla2016}. These rich phenomena have kept the QRM actively studied in fields such as quantum optics \cite{Fox2006} and interdisciplinary areas intersecting with condensed matter physics \cite{Zhang2021, Padilla2022}.

Extensions of the QRM, such as the asymmetric quantum Rabi model (AQRM, also known as biased QRM) \cite{Braak2011, Mao2018, Liu2017b}, also exhibit rich physical behavior. Theoretically, the introduction of an asymmetric term explicitly breaks the $\mathbb{Z}_2$ parity symmetry \cite{Braak2011}, lifting level crossings or degeneracies between adjacent energy levels, thereby suppressing the SPT in the ground state. However, when the asymmetry parameter takes specific integer or half-integer multiples of the bosonic mode frequency, level crossings reappear$-$a phenomenon attributed to a hidden symmetry \cite{Wakayama2017, Ashhab2020, Gardas2013, Mangazeev2021, Reyes-Bustos2021, Xie2022, Lu2022}. Yet, this hidden symmetry depends explicitly on system parameters and does not possess a simple form analogous to the parity operator in the standard QRM \cite{Ashhab2020, Xie2022, Lu2022}. Experimentally, the asymmetric term can be naturally implemented in circuit quantum electrodynamics (QED) systems \cite{Forn-Diaz2010, Yoshihara2018, Blais2021}, and its strength can be readily tuned externally.

Current research on the AQRM primarily focuses on analytical solutions and the search for a possible hidden symmetry operator. In existing literature, the energy splitting $\Delta$ of the two-level system is typically set around $1.0$, under which accidental degeneracies are observed to open and close at specific values of the asymmetry parameter \cite{Li2021, Zhong2014, Li2015, Batchelor2016, Guan2018}. To our knowledge, no studies have explored the regime where the ratio $\Delta/\omega$ is much larger than $1.0$ and the coupling strength $g > 1$, which corresponds to the superradiant phase region in the standard QRM. In this work, we investigate precisely the physics in this parameter regime.

In this paper, we set the energy splitting $\Delta = 10$. In the strong-coupling regime $(g > 1)$, energy level degeneracies are lifted when the asymmetry parameter deviates from $0.5$, $1.0$, $1.5$, etc., but are restored when the asymmetry parameter equals these values, though now occurring in excited states. Thus, even in the regime where $\Delta/\omega \gg 1$ and $g > 1$, the behavior remains similar to that at $\Delta \sim 1.0$ \cite{Li2021}, and the hidden symmetry persists. Using our previously developed method of two successive diagonalizations \cite{Yang2023}, we study the energy spectrum, effective potential, and wavefunctions of the model. The results indicate that matching energy levels in asymmetric double wells underlie this hidden symmetry, which is further corroborated by the wavefunctions of both ground and excited states.

\section{\label{sec:level2}Model and Method}
The Hamiltonian of the AQRM reads
\begin{equation}\label{H_AQRM}
H = \hbar\omega a^\dagger a + \frac{\Delta}{2}\sigma_x + g\sigma_z (a + a^\dagger) + \eta \sigma_z,
\end{equation}
where $a^\dagger (a)$ is creation (annihilation) operator of the single mode photon field with frequenc $\omega$, Pauli matrix $\sigma_x$ describes the two-level system with energy splitting $\Delta$, $\eta \sigma_z$ is the biased term, and g is the coupling strength. For convenience, we rescale the Hamiltonian by the mode frequency $\hbar\omega$, thus the parameters $\Delta$, $g$ and $\eta$ used in the following are dimensionless. In the case of $\eta=0$, the AQRM reduces to the standard QRM. Its Hamiltonian commutes with the parity operator $P=\sigma_x e^{i\pi a^\dagger a}$, satisfying $[H,P]=0$. Equivalently, the Hamiltonian remains invariant under the parity transformation $P^\dagger H P=H$, indicating that the standard QRM possesses a $\mathbb{Z}_2$ symmetry. As a result, parity serves as a good quantum number for labeling the eigenstates of the QRM. However, when a biased term is introduced $(\eta\neq 0)$, the Hamiltonian no longer commutes with the parity operator. Moreover, it is difficult to construct any simple and explicit parity-like operator that commutes with the Hamiltonian. This indicates that the $\mathbb{Z}_2$ symmetry is broken in the AQRM.

It is useful to use dimensionless position-momentum operators related to the creation (annihilation) operator by $a^\dagger = \frac{1}{\sqrt{2}}\left(\xi - \frac{\partial}{\partial \xi}\right)$ and $a = \frac{1}{\sqrt{2}}\left(\xi + \frac{\partial}{\partial \xi}\right)$ to rewrite the Hamiltonian as
\begin{eqnarray}\label{H_0}
&& H = H_0 + H_{\sigma},\\
&& \hspace{0.cm} H_0 = \frac{1}{2} \left(-\frac{\partial^2}{\partial\xi^2} + \xi^2\right),\\
&& \hspace{0.cm} H_{\sigma} = \frac{\Delta}{2}\sigma_x + \sqrt{2} g\sigma_z \xi+ \eta \sigma_z.
\end{eqnarray}
The matrix form of $H_\sigma$ is
\begin{equation}\label{Hsigma_matrix}
H_\sigma =
   \left(
  \begin{array}{cc}
    \sqrt{2} g \xi+\eta & \Delta/2 \\
    \Delta/2 & -\sqrt{2} g \xi-\eta \\
  \end{array}
\right).
\end{equation}
Thus Eq. \eqref{Hsigma_matrix} can be formally diagonalized and its eigenvales and eigenvectors read
\begin{eqnarray}
&& \epsilon_\pm(\xi) = \pm \frac{\Delta}{2} \sqrt{1 + \beta^2 \left(\xi+\frac{\eta}{\sqrt{2}g}\right)^2}, \label{eigenvalue}\\
&& \phi_{\pm}(\xi) = \frac{1}{\sqrt{2}} \left(\pm (1 \pm \gamma(\xi))^{\frac{1}{2}}, (1 \mp \gamma(\xi))^{\frac{1}{2}}\right)^T, \label{eigenvector}
\end{eqnarray}
where $\beta = \frac{2\sqrt{2} g}{\Delta}$ and $\gamma(\xi) = \frac{\beta \xi}{\sqrt{1 + \beta^2 \left(\xi+\frac{\eta}{\sqrt{2}g}\right)^2}}$. This finishes the first diagonization to solve the Schr\"odinger equation
\begin{equation}
H_\sigma \phi_\pm = \epsilon_\pm \phi_\pm. \label{sigma-diag}
\end{equation}
The second diagonalization is to solve the full Hamiltonian $H$ satisfying with the Schr\"odinger equation
\begin{equation}\label{Schrodinger}
H\Psi^E = (H_0 + H_\sigma)\Psi^E =  E\Psi^E.
\end{equation}
To proceed, it is useful to assume two complete basis $|\xi,\sigma\rangle :=|\xi\rangle|\sigma\rangle$ and $|\xi,\pm\rangle := |\xi\rangle|\pm\rangle$, and one can write the total wavefunction $\Psi^E$ as $\Phi^E(\xi,\sigma) = \langle \xi,\sigma|\Psi^E\rangle$ or $\psi^E_\pm(\xi) = \langle\xi,\pm|\Psi^E\rangle$. Utilizing the orthogonal basis $ 1 = \sum_\pm \int d\xi |\xi,\pm\rangle\langle\xi,\pm|$, the wavefunction can be written as $\Phi^E(\xi,\sigma)= \sum_\pm \phi_\pm(\xi) \psi^E_\pm(\xi)$.

\begin{figure*}
\centering
\includegraphics[width =1.5\columnwidth]{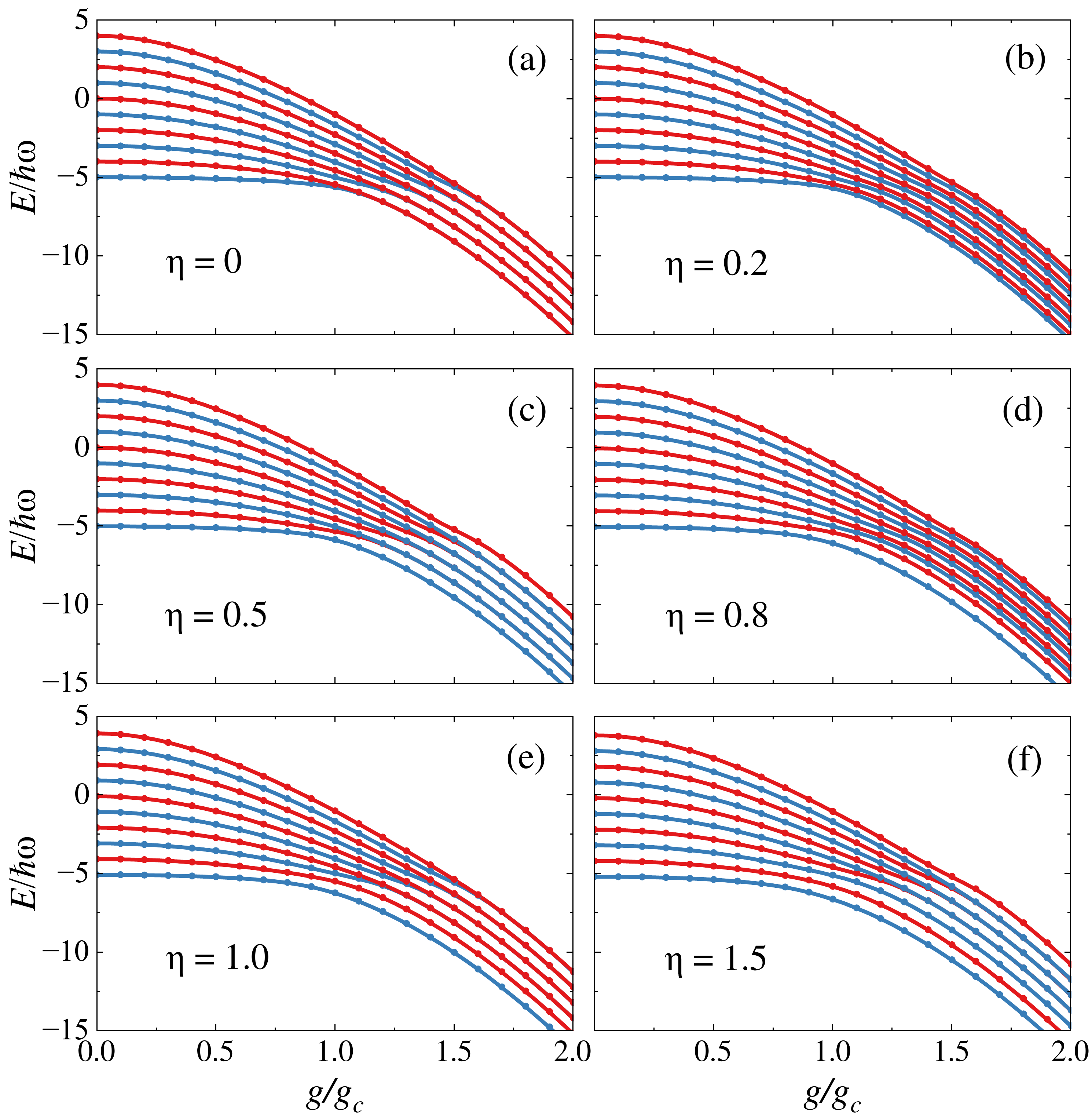}
\caption{Energy spectrum of the AQRM as functions of the coupling strength scaled by $g_c = \sqrt{1+\sqrt{1+\frac{\Delta^2}{16}}}$ \cite{Ying2015} with various asymmetric parameters (a) $\eta=0$, (b) $\eta=0.2$, (c) $\eta=0.5$, (d) $\eta=0.8$, (e) $\eta=1.0$, and (f) $\eta=1.5$. The lines are the results computed using our method and the solid dots are those obtained by numerical ED with the same parameter $\Delta = 10$,}
\label{fig1}
\end{figure*}

Then the Eq. (\ref{Schrodinger}) is written as 
\begin{equation}
\sum_\pm \left(H_0 + H_\sigma \right) \phi_\pm(\xi) \psi^E_\pm(\xi) = \sum_\pm E \phi_\pm(\xi) \psi^E_\pm(\xi), \label{Schrodinger1}
\end{equation}
multiplying from the left by $\phi_\pm^*$, we obtain
\begin{equation}
\langle \phi_\pm|H_0|\phi_\mp \rangle \psi^E_\mp(\xi) + \left(H_{0,\pm} + \epsilon_\pm \right)\psi^E_\pm(\xi) = E_\pm \psi^E_\pm(\xi). \label{Schrodinger2}
\end{equation}
The first term represents the coupling between $\phi_+$ and $\phi_-$. Considering the Born-Oppenheimer (B-O) approximation \cite{Born1927, Yang2023}, this coupling is neglected. The Eq. (\ref{Schrodinger2}) becomes
\begin{equation}
(H_{0,\pm} + \epsilon_\pm)\psi^E_\pm(\xi) = E_\pm \psi^E_\pm(\xi), \label{Schrodinger3}
\end{equation}
where $H_{0,\pm} = \phi^*_\pm H_0 \phi_\pm = H_0$. Eq. (\ref{Schrodinger3}) is the starting point of the following calculation.

In order to solve Eq. (\ref{Schrodinger3}), one inserts the complete basis $1 = \sum_n|n\rangle\langle n|$ of the standard harmonic oscillator into Eq. (\ref{Schrodinger3}) to obtain
\begin{eqnarray}
\sum_{m}\langle n|(H_0 + \epsilon_\pm)|m\rangle\langle m|\psi^E_\pm\rangle = E_{\pm,n} \langle n|\psi^E_\pm\rangle. \label{Schrodinger4}
\end{eqnarray}
In a truncated basis $|n\rangle, n = 0, 1, \cdots, N-1$, solving Eq. (\ref{Schrodinger4}) is equivalent to diagonalize the following $N\times N$ matrix
\begin{equation}\label{Htotal_matrix}
\left(
  \begin{array}{cccc}
    \langle 0|H_0+\epsilon_\pm|0\rangle & \cdots & \langle 0|\epsilon_\pm|N-1\rangle \\
    \vdots & \ddots & \vdots \\
    \langle N-1|\epsilon_\pm|0\rangle & \cdots & \langle N-1|H_0+\epsilon_\pm|N-1\rangle \\
  \end{array}
\right)
\end{equation}
Diagonalizing the above matrix, the eigenvales give the energy spectra of the AQRM. The total wavefunction of the model are expressed as
\begin{equation}\label{wavefunction}
\Psi^E_\sigma(\xi) =\phi_{\pm}(\xi) \sum_{n=1}^N a_n \varphi_n(\xi),
\end{equation}
where $a_n$ is the eigenvectors of the matrix (\ref{Htotal_matrix}), and $\varphi_n(\xi)$ is the wavefunctions of the standard harmonic oscillator. At this stage, the two successive diagonalizations are complete. 

We now examine the consequences of the B-O approximation. By neglecting the coupling between $\phi_+$ and $\phi_-$, the Hamiltonian decoples into two independent branches: the negative branch $H_0+\epsilon_-$ and the positive branch $H_0+\epsilon_+$. As illustrated in Appendix A of our published paper \cite{Yang2023}, the energy levels of the negative branch shift downward, while those of the positive branch move upward. A comparison with exact diagonalization reveals that the B-O approximation omits the avoided crossings between higher energy levels, which arise precisely from the coupling between $\phi_+$ and $\phi_-$. The onset of this coupling is governed by the two-level splitting $\Delta$. For $\Delta<1$, the B-O approximation fails, as it cannot capture the avoided crossings. In contrast, for $\Delta >1$, the B-O approximation yields reliable results for the lowest $\Delta$ energy levels. In the large $\Delta$ and large $g$ regime relevant to this work, the B-O approximation is particularly accurate. For this reason, we restrict our subsequent analysis to the negative branch of the Hamiltonian.

\section{\label{sec:level3} Main Results}
In this section, we employ the method introduced in Sec. \ref{sec:level2} to investigate the energy spectrum, effective potential, and wavefunctions of the AQRM, with comparisons to results from directly numerical exact diagonalization (ED). The obtained results elucidate the origin of the hidden symmetry and offer a clear physical picture. Finally, we briefly discuss the excited-state phase transitions in this model.

\subsection{\label{sec:level3A} Energy spectrum}

Fig. \ref{fig1} displays the energies of the ground state and low-lying excited states as functions of the coupling strength under different asymmetry parameters. To verify the accuracy of the present calculation, we also provide the results obtained via numerical ED for the same model parameter $\Delta = 10$, indicated by solid dots. It can be observed that our results exhibit excellent agreement with the exact values across both weak and strong coupling regimes. The following observations can be drawn from the figure:

(i) When the asymmetry parameter $\eta = 0$, the system corresponds to the standard QRM. In this case, starting from the ground state, every two adjacent energy levels become degenerate in the strong coupling regime$-$a direct consequence of the underlying $\mathbb{Z}_2$ symmetry.

(ii) For $\eta = 0.2$, all energy level degeneracies are lifted. However, when $\eta = 0.5$, degeneracy between adjacent levels is restored starting from the first excited state. As $\eta$ increases to $0.8$, all degeneracies are lifted again, but unlike the case at $\eta = 0.2$, the ground state and the first excited state are widely separated, while higher excited states (starting from the second excited state) exhibit small energy gaps, as shown in the comparison between Fig. \ref{fig1}(b) and (d). When $\eta$ further increases to $1.0$ and $1.5$, degeneracy reappears starting from the second and third excited states, respectively, as illustrated in panels (e) and (f).

(iii) Thus, at $\eta = 0.5, 1.0, 1.5, \ldots$, degeneracy between adjacent energy levels is restored beginning from the first, second, third excited states, etc. This phenomenon is attributed to the so-called hidden symmetry \cite{Wakayama2017, Ashhab2020, Gardas2013, Mangazeev2021, Reyes-Bustos2021, Xie2022, Lu2022}. In contrast, for general values of $\eta$ not equal to these specific values, all energy levels are split due to the breaking of $\mathbb{Z}_2$ symmetry by the asymmetric term. In the following, we will focus on discussing the origin of this hidden symmetry.

\begin{figure}[!htbp]
\centering
\includegraphics[width = \columnwidth]{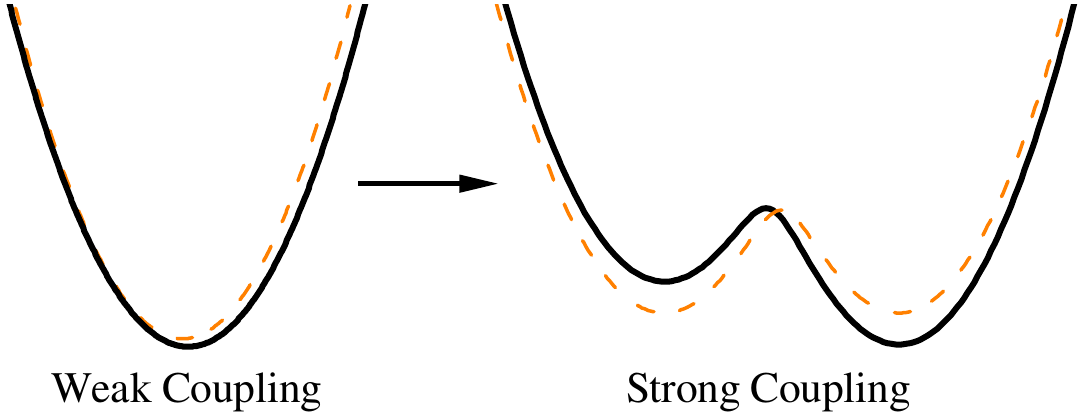}
\caption{Schematic diagram of the effective potential. The orange dashed curve represents the standard QRM ($\eta=0$), while the solid black curve corresponds to the AQRM ($\eta \neq 0$).}
\label{fig2}
\end{figure}

\begin{figure}[!htbp]
\centering
\includegraphics[width = \columnwidth]{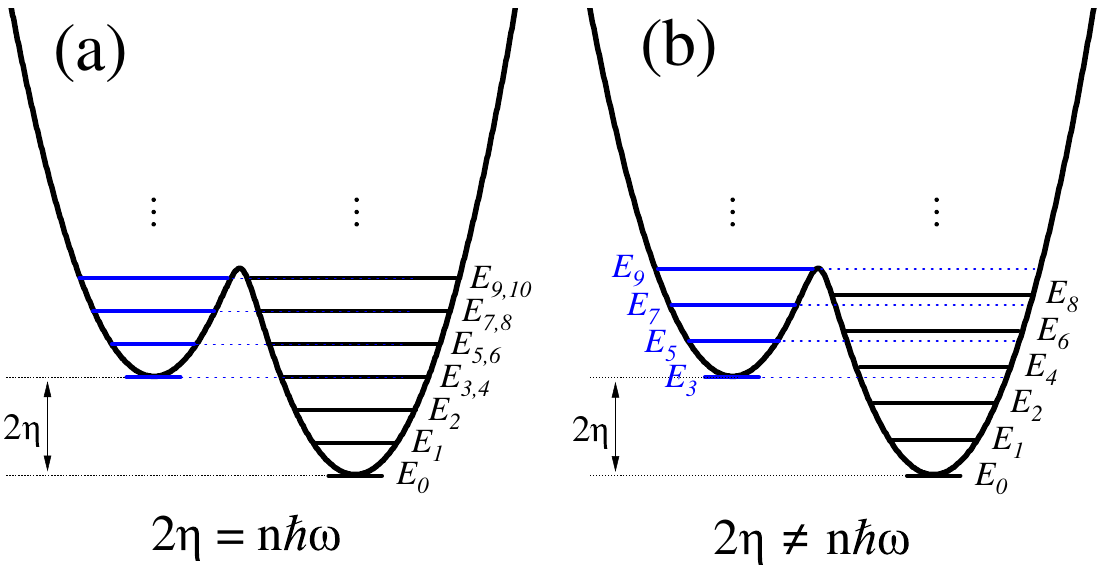}
\caption{Schematic diagram of energy-level matching in an asymmetric double-well potential. (a) $2\eta = n\hbar\omega$, the ground state in the higher well becomes degenerate with a certain excited state in the lower well; (b) $2\eta \neq n\hbar\omega$, all energy levels are non-degenerate and arranged in order of increasing energy.}
\label{fig3}
\end{figure}

\begin{figure*}[tbp]
\centering
\includegraphics[width =1.5\columnwidth]{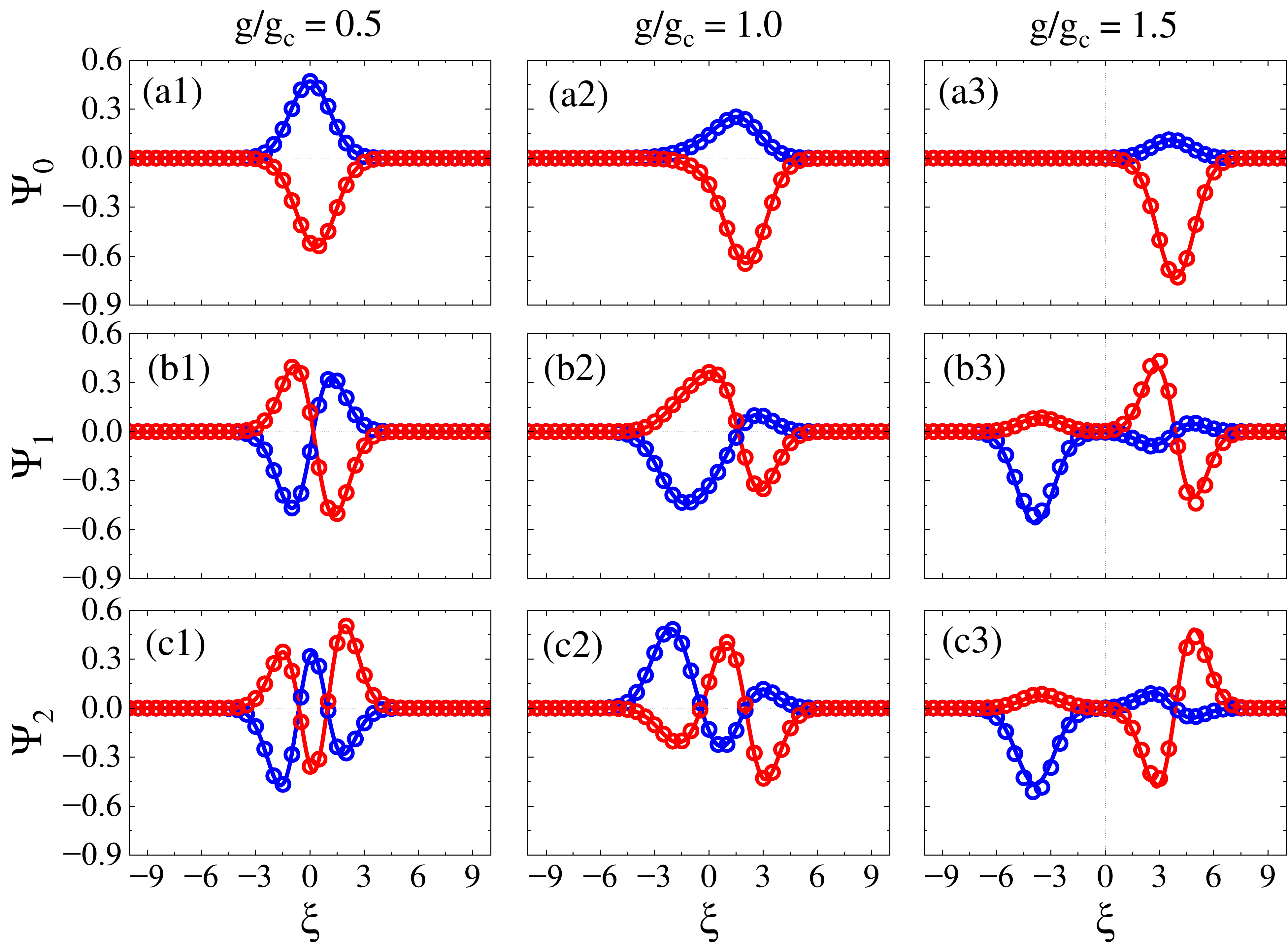}
\caption{Wavefunctions of the AQRM at $\eta=0.5$. The first, second, and third rows correspond to the ground state, first excited state, and second excited state, respectively. Each column represents a different coupling strength scaled by $g_c = \sqrt{1+\sqrt{1+\Delta^2/16}}$. Solid lines denote the results obtained by our method and circles represent those obtained from ED. Red and blue curves indicate the spin-up and spin-down components of the wavefunctions.}
\label{fig4}
\end{figure*}

\begin{figure*}[tbp]
\centering
\includegraphics[width = 1.5\columnwidth]{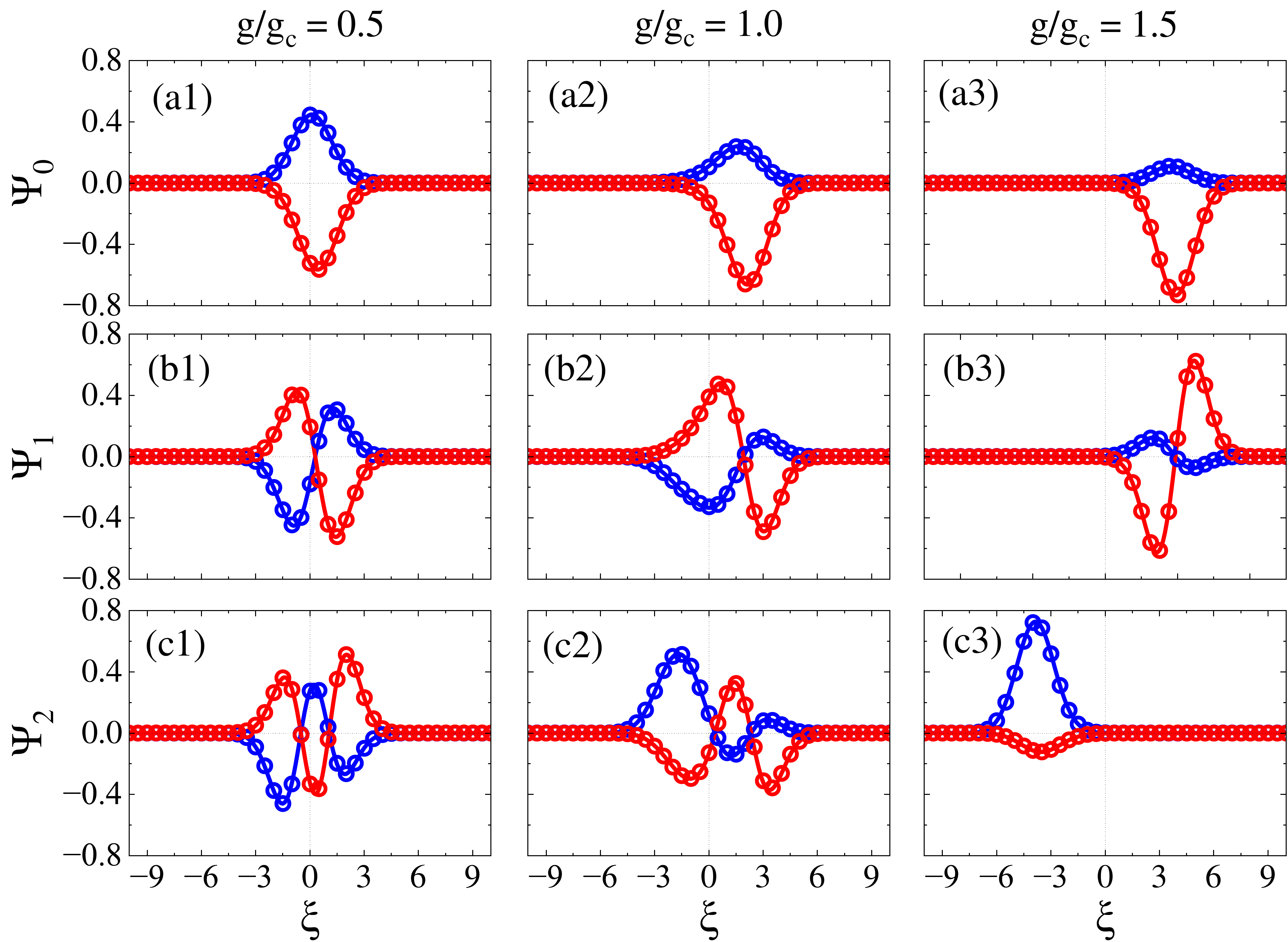}
\caption{Wavefunctions of the AQRM at $\eta=0.8$. Other information is the same as Fig. \ref{fig4}.}
\label{fig5}
\end{figure*}

\subsection{\label{sec:level3B} Effective Potential}

In Sec. \ref{sec:level2}, we obtained the expressions for $\epsilon_{\pm}(\xi)$, and here, we consider only the negative branch. Combined with $H_0$, the effective potential of the system can be expressed as follows,
\begin{equation}\label{Veff0}
V_{\text{eff}}(\xi) = \frac{1}{2} \xi^2 - \frac{\Delta}{2} \sqrt{1 + \beta^2 \left(\xi + \frac{\eta}{\sqrt{2}g}\right)^2}.
\end{equation}
Expanding this effective potential as a Taylor series yields, 
\begin{equation}\label{Veff1}
V_{\text{eff}}(\xi) = C_0 + C_1 \xi + C_2 \xi^2 + C_3 \xi^3 + C_4 \xi^4 + O[\xi]^5,
\end{equation}
where the coefficients $C_0$, $C_1$, $C_2$, $C_3$, and $C_4$ are functions of the parameters $\Delta$, $\eta$, and $g$, reading
\begin{eqnarray}
&& C_0 = -\frac{\Delta}{2\sqrt{2}g}\left(2g^2+\beta^2\eta^2\right)^{1/2},\\
&& \hspace{0.cm} C_1 = -\frac{4g^2\eta}{\Delta}\left(2g^2+\beta^2\eta^2\right)^{-1/2},\\
&& \hspace{0.cm} C_2 = \frac{1}{2}-\frac{4\sqrt{2}g^5}{\Delta}\left(2g^2+\beta^2\eta^2\right)^{-3/2},\\
&& \hspace{0.cm} C_3 =\frac{64g^8\eta}{\Delta^3}\left(2g^2+\beta^2\eta^2\right)^{-5/2},\\
&& \hspace{0.cm} C_4 =\left(\frac{32\sqrt{2}g^{11}}{\Delta^3}-\frac{512\sqrt{2}g^{11}\eta^2}{\Delta^5}\right)\left(2g^2+\beta^2\eta^2\right)^{-7/2}.
\end{eqnarray}
Obviously, when $\eta = 0$, all odd-powered terms in $\xi$ vanish, corresponding to the standard QRM. In this case, the effective potential can be approximately regarded as a standard harmonic oscillator potential in the weak-coupling regime, with its minimum located at $\xi = 0$. In the strong-coupling regime, the potential evolves into a symmetric double-well structure, where the original minimum becomes a local maximum that remains at $\xi = 0$, as illustrated by the orange dashed curve in Fig. \ref{fig2}. For $\eta \neq 0$, the odd-powered terms in $\xi$ are present. In the weak-coupling regime, the effective potential still resembles a harmonic oscillator potential, but its minimum is shifted. Significant changes occur in the strong-coupling regime: the effective potential becomes asymmetric, forming an asymmetric double-well structure, and the position of the local maximum is also displaced, as depicted by the solid black curve in Fig. \ref{fig2}. Intriguing physics arises in this asymmetric double-well.

Fig. \ref{fig3} illustrates a schematic diagram of energy level matching in the asymmetric double-well potential. In the strong coupling regime, each well of the asymmetric double-well effectively behaves as a harmonic oscillator with frequency $\omega$, and their minima are offset by $2\eta$. The energy levels are sketched within the effective potential, ignoring the zero-point energy for clarity. As shown in (a), when the condition $2\eta = n\hbar\omega$ is satisfied, the ground state in the higher well becomes degenerate with a certain excited state in the lower well, while the ground state and several low-lying excitations in the lower well remain non-degenerate. A larger value of $\eta$ results in more non-degenerate levels in the lower well, as can be readily observed in (a). Conversely, if $2\eta \neq n\hbar\omega$, all energy levels are non-degenerate and arranged strictly by energy, as depicted in (b). This intuitive picture explains the phenomena observed in the previous subsection and demonstrates that, in the regime where $\Delta/\omega \gg 1$ and $g > 1$, energy-level matching within the asymmetric double-well serves as the origin of the hidden symmetry. This interpretation is further supported by wavefunctions of the AQRM, as discussed in the following subsection.

\subsection{\label{sec:level3C} Wavefunctions}

Figs. \ref{fig4} and \ref{fig5} display the wavefunctions of the AQRM under different asymmetry parameters. The first, second, and third rows correspond to the ground state, first excited state, and second excited state, respectively, while each column represents a different coupling strength. Solid lines denote the results computed using our method, and circles represent those obtained from directly numerical ED; the two remain in excellent agreement across the entire coupling regime. Red and blue curves indicate the spin-up and spin-down components of the wavefunctions.

Fig. \ref{fig4} shows the wavefunctions at $\eta=0.5$. As indicated in Fig. \ref{fig1}(c), under this parameter, the ground state is non-degenerate, while adjacent energy levels become degenerate from the first excited state onward in the strong-coupling regime. The following observations can be made:

(i) At $g/g_c=0.5$, the effective potential slightly deviates from the standard harmonic oscillator, and the wavefunctions remain largely similar, as shown in the first column;

(ii) At $g/g_c=1.0$, although the asymmetric double-well has not fully formed, the wavefunctions are shifted due to the presence of the biased term, as illustrated in the second column;

(iii) Interesting behavior emerges in the strong-coupling regime. At $g/g_c=1.5$, the ground-state wavefunction resembles that of the harmonic oscillator ground state, albeit shifted to the right [panel (a3)]. Notably, the first excited-state wavefunction exhibits left and right components corresponding to the ground and first excited states of the harmonic oscillator, respectively [panel (b3)]. A similar wavefunction is observed for the second excited state [panel (c3)];

(iv) The third and fourth excited states (not shown here) further follow this trend: their left and right segments correspond to the first and second excited states of the harmonic oscillator. Higher excited states adhere to the same regularity. These findings fully corroborate the physical picture of energy-level matching within the effective potential, as proposed in the previous subsection.

Fig. \ref{fig5} presents the wavefunctions for $\eta=0.8$. According to Fig. \ref{fig1}(d), all energy levels are non-degenerate under this condition. The ground state and the first excited state reside in the lower well, corresponding to the ground and first excited states of that well, as shown in panels (a3) and (b3), respectively. The second excited state occupies the higher well and corresponds to the ground state of that well, as depicted in panel (c3). These results are fully consistent with the physical picture described earlier.

\begin{figure}[bp]
\centering
\includegraphics[width =0.8 \columnwidth]{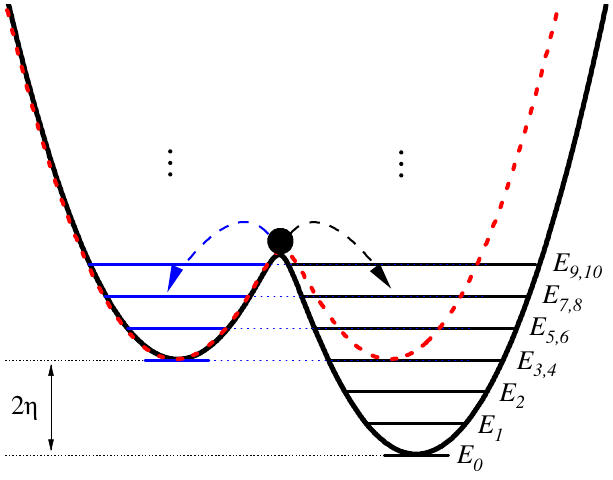}
\caption{Schematic of the effective potential. The solid black curve corresponds to the asymmetric double-well potential in the AQRM, while the red dashed curve depicts a fictitious effective symmetric double-well potential, which produces energy level degeneracies.}
\label{fig6}
\end{figure}

\subsection{\label{sec:level3D} Excited-state Quantum Phase Transition}
At specific values of the asymmetry parameter, the degeneracy observed in high-lying excited states indicates the existence of an excited-state quantum phase transition in the AQRM, a phenomenon that has not been reported in previous literature. This transition can be intuitively understood through the schematic illustration of the effective potential in Fig. \ref{fig6}, where the solid black curve represents the asymmetric double-well, and the red dashed curve depicts a fictitious effective symmetric double-well, which can produce energy level degeneracies, as shown in Fig. \ref{fig6}. This fictitious symmetric double-well suggests the restoration of a parity-like symmetry in the system, i.e., the hidden symmetry. In the strong-coupling regime, the excited-state quantum phase transition should occur due to the breaking of this hidden symmetry, while in the weak-coupling regime, the hidden symmetry is restored, corresponding to the normal phase of the system.

A quantitative characterization of this possible excited-state quantum phase transition$-$including the construction of an effective model, computation of critical exponents, and analysis of scaling behavior$-$is left for future investigation.

\section{\label{Conclusion}Conclusion}
We investigate the AQRM using the method of two successive diagonalizations, with a focus on the parameter regime where $\Delta/\omega \gg 1$. When the asymmetry parameter takes specific values, degeneracies emerge among the excited-state energy levels, indicating the presence of a hidden symmetry. We find that the origin of this hidden symmetry lies in the matching of energy levels within the asymmetric double-well potential, a conclusion further supported by the wavefunctions of both the ground and excited states. Finally, we briefly discuss the excited-state quantum phase transition and its relation with the hidden symmetry in this model.

\section{Acknowledgments}
The work is partly supported by the National Key Research and Development Program of China (Grant No. 2022YFA1402704) and the programs for the National Natural Science Foundation of China (Grant No. 12247101).




%

\end{document}